 \documentclass [12pt,a4paper      ]{article}
\usepackage{graphics}
\usepackage{times}

\DeclareFontFamily{OT1}{times}{}
\DeclareFontShape {OT1}{times}{m }{n }{ <-> ptmr }{}
\DeclareFontShape {OT1}{times}{bx}{n }{ <-> ptmb }{}
\DeclareFontShape {OT1}{times}{m }{it}{ <-> ptmri}{}
\DeclareFontShape {OT1}{times}{bx}{it}{ <-> ptmbi}{}
\usepackage{amsmath}
\usepackage{amsfonts}
\usepackage{amssymb}
\usepackage{latexsym}
\newcommand{\cl}{C \kern -0.1em \ell} % Clifford algebra
\newcommand{\VEC}{\vec{\kern +.1em[}} % VEC-tor left  bracket   |>.. 
\newcommand{\TOR}{\vec{\kern +.2em]}} % vec-TOR right bracket   ..>| 
\newcommand{\BRA}{\langle\kern -.2em\langle} % Dirac BRA      <<...| 
\newcommand{\KET}{\rangle\kern -.2em\rangle} % Dirac KET      |...>> 

\numberwithin{equation}{section}
\begin{document}

\title{\bf\vspace{-2.5cm} Dirac's Wave Mechanical Theory of the Electron
                          and Its Field-Theoretical
                          Interpretation\footnote{\emph{Editorial note:}
                          Published in
                          Physikalische Zeits. {\bf 31} (1930) 120--130,
                          reprinted and translated in~\cite{LAN1930A}.
                          This is Nb.~4 in a series of four papers
                          on relativistic quantum mechanics
                          \cite{LAN1929B,LAN1929C,LAN1929D,LAN1930A}
                          which are extensively discussed
                          in a commentary by Andre Gsponer
                          and Jean-Pierre Hurni \cite{GSPON1998B}.
                          Initial translation by J\'osef Illy and 
                          Judith Konst\'ag Mask\'o. Final translation
                          and editorial notes by Andre Gsponer.}}
%                         ======================================

\author{By Cornel Lanczos in Berlin\\
       (Received on October 31, 1929)}

\date{Version ISRI-04-13.3 ~~ \today}

\maketitle

\begin{abstract}

\emph{Editorial note:} In this lecture, held at the meeting of the Berlin Physical Society, October 25, 1929, Lanczos discusses a nonlinear generalization of Proca's equation (derived from his ``fundamental equation'') such that the mass is multiplied by a scalar field function.

\end{abstract}

The admirable success of recent theoretical research in the field of atomic physics, on the one hand, and the undeniable and irreplaceable wealth of great field theoretical achievements of the last century, on the other, have accustomed us, in recent years, more and more to a type of unpleasant dualistic thinking which is apparently unavoidable. As a result of this dualistic thinking, we must assume entirely different structures for the laws governing phenomena in empty space --- in the aether, in the ``field'' as we like to say --- and for the laws of phenomena in which interactions between atoms and molecules, or also between radiation and matter, play a role. The difference is not one of degree, but is so incisive that one needs a completely different intellectual attitude for conceptually understanding one or the other types of contexts.

   There have been no lack of attempts by this or that camp at abolishing these differences in favor of a unified point of view. Attempts have been made to consider the field theoretical ideas as belonging to a somewhat childish stage of physical thinking. These ideas would have to be sacrificed with the advancement of knowledge in favor of less intuitive concepts and formulations, which are considerably more complicated, almost dialectical, though [they are] in better agreement with immediate experience. On the other hand, even one of the greatest of the discoverers in the new field, Schr\"odinger, has made an attempt at putting the quantum laws on a pure field theoretical footing by constructing a kind of wave optics based on analytical dynamics and Fermat's principle. The function, which he denotes by $\psi$, which ought to govern the entire atomic realm as far as quantum-like processes are concerned, is distributed over the entire space and has a completely field theoretical character. Moreover, he gave this function a direct physical interpretation by having brought it into direct connection with the distribution of electric charge --- with the density of electricity.  This approach has, however, gained little credibility for two reasons. First, contrary to early expectations, it is impossible to construct an electron by the interference of spatially dispersing waves so as to arrive at a formation as stable as one would like to have for the electron in an empirically founded field theory. Furthermore, the entire theory is founded on something foreign to the field theoretical approach, namely, on analytical dynamics. Though the rule of how to arrive at a wave equation does lead to a wave theory in the usual space-time world in the case of a single body, this result, nonetheless, is better considered as one of mere chance. Whenever several bodies act on each other, it becomes senseless to try to connect the wave equation formed according to the corresponding rule --- which has stood the test of experience  --- with reasonable field theoretical ideas.

   However great the success which has been achieved by the wave equation found by Schr\"odinger, of late years a gap has evolved between the authority of analytical dynamics and the authority of the fundamental wave equation deduced from analytical dynamics in a particular way.

   From the very beginnings of the new abstract quantum physics that superseded Bohr's atom model, a peculiar difficulty arose, the essence of which could not be so easily explained. I have in mind what is called, in short, the ``rotating electron'' or ``electron spin.'' It is well-known how fundamentally this hypothesis controls the entire system of [spectroscopic] terms; what is more, the whole periodic system --- indeed, the entire shell structure of the elements --- has found its place so that by means of this hypothesis, which was introduced somewhat $\emph{post festum}$, even the old Bohr model would have been able to describe facts qualitatively correctly in a quantitatively sketchy manner.

   Such a hypothesis could evidently be incorporated into quantum mechanics by means of a suitable reinterpretation of concepts. But all was so artificial, so out of the clouds, so incomprehensible that one could hardly avoid the uneasy feeling, or brush off the suspicion, that something fundamental was lacking or that there was something wrong there. And that, in fact, was the case.

   Thanks to the young English theoretician, Dirac, the solution of the enigma was found. He did not brood at all over the electron spin; this welcome result came to light rather later as an unexpected --- but just for that very reason an even more precious gift --- as so often happens with scientific discoveries.

   It was another matter that made him reconsider Schr\"odinger's theory. The Schr\"odinger equation had a remarkable feature: it was first-order in time, but second-order in spatial coordinates. How could it happen, since the theory of relativity had established so convincingly the essential equality of space and time? Either everything must be of first-order, or everything must be of second-order. What is more, both could be conceived to be valid at the same time. It might be that a first-order equation is the primary, and then, by another differentiation, a second-order equation (now also second-order in time) of a kind from which the Schr\"odinger wave equation could be obtained.

   Does such a first-order wave equation exist?  If it existed, it would have been discovered long ago, since a fundamental importance had always been attributed to it from Laplace on --- remember also its special case, the potential equation --- in mathematics and in the entire field of theoretical physics. But Dirac, following his own way since the foundation of the new quantum theory, could do many things which until then where unsuspected, because he was working with quantities which had not been in common use earlier.

   It is a commonplace truth for us that two times three is as many as three times two. Mathematics, however, knows of quantities for which the interchangeability of the order of factors fails. These are the so-called non-commuting quantities. And Dirac had the original idea to track the failure of earlier physics, in relation to the subtleties of atomic dynamics, back to the practice of always calculating as if the commutation law were valid. Indeed, nature realizes quantities that do not follow this rule and for which precisely another commutation law is valid in which Planck's quantum of action $h$, so fundamental for all quantum phenomena, enters in a characteristic way. He calls such non-commuting quantities ``$q$-numbers'' in contrast to usual numbers which he calls ``$c$-numbers.'' By using such $q$-numbers, it was not difficult for him to find the required equation. It contained, indeed, only first derivatives, both of factors and unknown quantities entered as $q$-numbers. Furthermore, a repeated application of the equation provided the already-known Schr\"odinger equation as a mathematical consequence. Hence, this equation is here not the starting point but rather a consequence. And, in addition, the new conception proved its superiority. When an external field was present, one did obtain exactly the Schr\"odinger equation. Additional terms were left over, and a closer analysis revealed that these terms provided just the corrections for the spectral terms that had to be forced from the unnatural and, in its quantitative details, \emph{ad hoc} hypothesis of the spinning electron. It was a remarkable success, not to be attributed to a mere accident, considering the entirely different viewpoint of the deduction.

   However much of Dirac's conception of quantum mechanics differs from the other ways adopted, it coincides with those in its consequences because Heisenberg's fundamentally new method of using matrices --- i.e., quadratic arrangements of numbers --- in the discussion goes parallel to that of Dirac in that these matrices do not follow the law of commutativity of multiplication. One can regard a matrix as a direct representation of a ``$q$-number.''  One can get from Heisenberg to Dirac by regarding matrices consisting of a great number, even of infinitely many elements in general, in totality as a whole, by labeling it with a letter and introducing it in mathematics as a non-commuting algebraic number. Conversely, one can avoid Dirac's non-commuting quantities by replacing them with matrices that are then built up of normal numbers, which corresponds more to the usual way of thinking.

   If we do so with Dirac's wave equation, the use of matrices leads to the effect that in place of a single equation we shall have a system of equations and in place of the only fundamental function $\psi$ we shall have a series of functions of that kind. The one Dirac equation becomes a system of first-order partial differential equations. To put it more clearly, a system of four complex equations with four complex unknowns.

   The development from Schr\"odinger to Dirac may be illustrated by an historical analogy. It corresponds perfectly to the development of wave optics from Fresnel to Maxwell. Fresnel used a wave equation which was a single scalar equation. Though it was known that light must be a vectorial phenomenon because of the transversality of light vibration, a detailed conception was left out of consideration until it appeared in Maxwell's electromagnetic theory by itself. Though Maxwell's equations are of the first-order, the propagation of electromagnetic phenomena with light velocity and the wave equation can be derived from them, all having been taken by Fresnel as a basis of his theory. Indeed, the analogy with quantum mechanics is complete. The difficulty the Schr\"odinger scalar theory ran into because of the spin effect may be compared to the difficulty Fresnel met by letting light through a boundary layer between two media with different optical densities. Fresnel hinted at the correct relations with an admirable sense of tact, but was compelled to make assumptions which have proved to be problematic. Maxwell's theory provided the right boundary conditions automatically and with necessity.

   In one respect, our analogy is yet not really correct. Maxwell's theory was a true field theory in the sense that one could ascribe to the quantities in it a rational meaning, representable by geometric, i.e., vectorial formations. By comparison, let us examine Dirac's equations. They can be put in the following form: \footnote{Cf., e.g., H. Weyl, Gruppentheorie und Quantummechanik, (Hirzel Leipzig, 1928) p.~171. Dirac's original works were published in the Proc. Roy. Soc., particularly {\bf 117}, 610; {\bf 118}, 351, 1928.}
\begin{equation*}\label{1}
\left.
\begin{aligned}
\frac{1}{i c} \frac{\partial \psi_3}{\partial t} +i \frac{\partial \psi_4}{\partial x} + \frac{\partial \psi_4}{\partial y} + i \frac{\partial \psi_3}{\partial z} + \frac{2\pi m c}{h} \psi_1 &= 0  \\
\frac{1}{i c} \frac{\partial \psi_4}{\partial t} +i \frac{\partial \psi_3}{\partial x} - \frac{\partial \psi_3}{\partial y} - i \frac{\partial \psi_4}{\partial z} + \frac{2\pi m c}{h} \psi_2 &= 0  \\
\frac{1}{i c} \frac{\partial \psi_1}{\partial t} -i \frac{\partial \psi_4}{\partial x} - \frac{\partial \psi_2}{\partial y} - i \frac{\partial \psi_1}{\partial z} + \frac{2\pi m c}{h} \psi_3 &= 0  \\
\frac{1}{i c} \frac{\partial \psi_2}{\partial t} -i \frac{\partial \psi_1}{\partial x} + \frac{\partial \psi_1}{\partial y} + i \frac{\partial \psi_2}{\partial z} + \frac{2\pi m c}{h} \psi_4 &= 0
\end{aligned} \quad
\right\}
\tag{1}
\end{equation*}
($m$ is the mass of the electron, $c$ is the velocity of light, $h$ is the quantum of action). The four complex quantities $\psi_1$  to $\psi_4$  are to be regarded as functions of the three spatial coordinates $x, y, z$ and the time $t$.

   The equations appear rather unattractive. The various coordinates enter in very different ways. So the terms differentiated with respect to $x$ and $z$ are multiplied everywhere by $i$, while this is not the case with $y$. One thus has the impression that a vectorial framework would be given in space with numbered axes. In fact, it is not so. We may carry out an arbitrary rotation of the axes, and even more far-reaching, an arbitrary Lorentz transformation without any change of the system of equations. As we say, there is an invariance under the group of Lorentz transformations. This also corresponds completely to the character of Maxwell's equations.

   But the laws the complex quantities $\psi$ follow under transformation have a structure radically different from what we are accustomed to for a vector or a field intensity and the like. We ought to vaguely imagine these $\psi$ functions as auxiliary quantities of fundamental importance for the events of quantum mechanics but without the possibility of ascribing to them a direct meaning. Fortunately --- and this is very remarkable --- one can construct quadratic expressions of the $\psi$ quantities that have a proper field theoretic meaning. Let us write out the following four expressions:
\begin{equation*}\label{2}
\left.
\begin{aligned}
  (&\psi_1 \psi_2^* + \psi_2 \psi_1^*) - (\psi_3 \psi_4^* + \psi_4 \psi_3^*)  \\
i[(&\psi_1 \psi_2^* - \psi_2 \psi_1^*) - (\psi_3 \psi_4^* - \psi_4 \psi_3^*)] \\
  (&\psi_1 \psi_1^* - \psi_2 \psi_2^*) - (\psi_3 \psi_3^* - \psi_4 \psi_4^*)  \\
i[(&\psi_1 \psi_1^* + \psi_2 \psi_2^*) + (\psi_3 \psi_3^* + \psi_4 \psi_4^*)]
\end{aligned}  \quad
\right\}
\tag{2}
\end{equation*}
(the asterisks denote conjugate complex quantities). It can be seen that these four quantities together make a so-called ``four-vector'' that unites what occurs in the usual three-dimensional interpretation as current density (the first three expressions) and charge density (the fourth expression, without the factor $i$). The fourth expression corresponds to a quantity interpreted by Schr\"odinger as ``the density of electricity.'' In this interpretation, vectors with components of the first three rows would complete electric density by electric current. Indeed, as a consequence of field equations \eqref{1} connecting the quantities $\psi$, the important circumstance exists that a \emph{conservation law} holds: the increase of electricity in an arbitrary volume is equal to the amount of electricity introduced by the current.

   For reasons discussed above, we are inclined to give the quantities introduced by Schr\"odinger a more abstract meaning by interpreting them as statistical quantities, as \emph{probabilities}, namely, as the number of particles found in unit volume on the average. The completing vector then means the ``probability flux:'' the number of particles crossing the unit surface per unit time on the average. The conservation law mentioned means here the conservation of the number of particles endowed with fixed charge.

   So much for the essentials of Dirac's theory. And now I arrive at the point where my own investigations come in. I have made the attempt --- better to say a chance [gelegentliche] observation led me --- to enlarge the [Dirac's] theory in such a way that it obtains a true field theoretic character. We cannot yet do it actually with the system of equations given above. There is, however, in the last term a constant ${2\pi m c}/{h}$ that I denoted by $\alpha$. It is only its square that enters the wave equation and we are not sure whether to take the last term with positive or negative sign --- both are permitted. Now the idea comes to take into consideration not this or that possibility but both together. Evidently we must rather ascribe each sign to a separate system of functions $\psi$ instead of to one and the same one. We shall thus have 8 complex equations with 8 complex functions $\psi$. The entire system, now equivalent to two equations of Dirac (one with $+\alpha$, one with $-\alpha$), has more general transformation properties than any of the equations taken alone, and it is now possible to implement the longed-for field theoretical approach. The mathematical development impressively shows how unambiguously the way is prescribed to the realization of the program, but I shall leave it aside, and shall just give the result.\footnote{The original works of the author were published in the Zeits.\ f.\ Phys., {\bf 57}, 447, 474, 484, 1929. (\emph{Editorial note:} See Refs.~\cite{LAN1929B,LAN1929C,LAN1929D}.)}  The foregone historical analogy used with Maxwell's equations will now provide a direct factual background.

   Let us recall the well-known Maxwell equations, the fundamental equations of the electromagnetic field. They read in vacuum:
\begin{equation*}\label{3}
\begin{array}{c}
\left.
\begin{array}{c}
\dfrac{1}{c} \dfrac{\partial \mathcal{E}}{\partial t} - \text{rot} \, \mathcal{H} = 0 \\
\text{~~~ ~~~ ~~~ \, div} \, \mathcal{E} = 0
\end{array}
\right\} \, (A) \\
\left.
\begin{array}{c}
\dfrac{1}{c} \dfrac{\partial \mathcal{H}}{\partial t} + \text{rot} \, \mathcal{E} = 0 \\
\text{~~~ ~~~ ~~~ div} \, \mathcal{H} = 0 
\end{array}
\right\} \, (B) 
\end{array}
\tag{3}
\end{equation*}
For the sake of a better understanding, we use the notation of elementary vector analysis and disregard the tensor analytic formulation offered by the theory of relativity, which made possible the first great general view.

   As is well-known, Maxwell carried out various modifications on his equations in order to apply them to ponderable bodies. However, the theory of electrons recognized a part of these modifications superfluous and retained only the electron charge and the corresponding convection current. The equations with this addition read:
\begin{equation*}\label{4}
\begin{array}{c}
\left.
\begin{array}{c}
\dfrac{1}{c} \dfrac{\partial \mathcal{E}}{\partial t} - \text{rot} \, \mathcal{H} = -\dfrac{4\pi \rho}{c} v \\
\text{~~~ ~~~ div} \, \mathcal{E} = 4 \pi \rho
\end{array}
\right\} \, (A) \\
\left.
\begin{array}{c}
\dfrac{1}{c} \dfrac{\partial \mathcal{H}}{\partial t} + \text{rot} \, \mathcal{E} = 0 ~~~ ~~~ ~~~ \\
\text{~~~ ~~~ div} \, \mathcal{H} = 0 ~~~ ~~
\end{array}
\right\} \, (B) 
\end{array}
\tag{4}
\end{equation*}
The system of equations (\ref{4}B) may be solved by introducing potentials, the ``vector potential'' $\mathcal{A}$, and the ``scalar potential'' $\varphi$ (combined together in four-dimensional union in a ``four-potential'' $\varphi_i$, also called ``vector potential,'' so that we shall call \, $\mathcal{A}$ ``spatial vector potential''):
\begin{equation*}\label{5}
\left.
\begin{aligned}
\mathcal{H} &= \text{rot} \, \mathcal{A} , \\
\mathcal{E} &= -\frac{1}{c} \frac{\partial \mathcal{A}}{\partial t} - \text{grad} \, \varphi
\end{aligned}
\right\}
\tag{5}
\end{equation*}
(As it was shown by the theory of relativity, the two expressions for $\mathcal{H}$ and $\mathcal{E}$ definitely belong together, because they represent only one object in four dimensions, expressed in three dimensions by the equation $\mathcal{H}= \text{rot} \, \mathcal{A}$.)

   If we consider Dirac's equations doubled as indicated above, we can also unite them by a convenient arrangement in a system very similar to that of Maxwell.

   The only difference is that in Maxwell's equations only the electromagnetic field intensity plays the role of a fundamental field quantity. The vector potential is introduced only as an auxiliary quantity to help in solving the equations. In our new system, however, the vector potential together with the field strengths are field quantities; in the equations they occur as primary quantities. The second system of Maxwell's equations and equations \eqref{5} are equivalent; we could replace equations (\ref{4}B) by equations \eqref{5} from the very beginning. We do not do so since equations \eqref{5} contain no direct relations between field intensities, but introduce new entities. Here, however, we have reasons to replace the second system of Maxwell by equations \eqref{5}. In effect, the first of the Maxwellian system occurs now in a modified form in which the potentials also appear:
\begin{equation*}
\left.
\begin{aligned}
\frac{1}{c} \frac{\partial \mathcal{E}}{\partial t} - \text{rot} \, \mathcal{H} &= \alpha^2 \mathcal{A} \\
\text{div} \, \mathcal{E} &= -\alpha^2 \varphi
\end{aligned}
\right\} 
\notag
\end{equation*}
The second system of Maxwell's equations, i.e., the connection between field intensity and vector potential, is left unchanged.

   Summarizing, we have ten field quantities: the electromagnetic field intensity (6 components) and the vector potential (4 components).  Ten field equations hold between them. We shall write them once more in a collected form:\footnote{\emph{Editorial note:} These equations are actually Proca's equations written in elementary vector form.  Just like Proca in 1936, Lanczos in 1929 did not realize that these equations do not apply to a spin~1/2 electron field, but rather to a massive spin~1 field.  For more details, see section~11 in~\cite{GSPON1998B}.}
\begin{equation*}\label{6}
\begin{array}{c}
\left.
\begin{array}{c}
\dfrac{1}{c} \dfrac{\partial \mathcal{E}}{\partial t} - \text{rot} \, \mathcal{H} = \alpha^2 \mathcal{A} \\
\text{~~~ ~~~ ~~~ ~~~ div} \, \mathcal{E} = -\alpha^2 \varphi
\end{array}
\right\} \, (A) \\
\left.
\begin{array}{c}
\mathcal{E} = -\dfrac{1}{c} \dfrac{\partial \mathcal{A}}{\partial t} - \text{grad} \, \varphi \\
\mathcal{H} = \text{rot} \, \mathcal{A} ~~~ ~~~ ~~~ ~~~ ~~~ ~~~ ~~
\end{array}
\right\} \, (B) 
\end{array}
\tag{6}
\end{equation*}
Because of equations  (\ref{6}A) a connection holds between $\mathcal{A}$ and $\varphi$, which for simplicity is assumed to hold also in Maxwell's theory, but here this connection appears of necessity, i.e.:
\begin{equation*}\label{7}
\text{div} \, \mathcal{A} + \frac{1}{c} \frac{\partial \varphi}{\partial t}= 0 .
\tag{7}
\end{equation*}
It may be added to our system as the 11th equation and represents the equation of continuity of electricity.

   Indeed, if we compare system \eqref{6} and equations \eqref{4} of the theory of electrons, we see that the assertion of Dirac's equations is to be understood simply as saying that the four-current is proportional to the four-potential:
\begin{equation*}\label{8}
\left.
\begin{aligned}
i = \rho v &= - \frac{c}{4 \pi} \alpha^2 \mathcal{A} \\
\rho &= -\frac{1}{4\pi} \alpha^2 \varphi
\end{aligned}
\right\}
\tag{8}
\end{equation*}
Equation \eqref{7} contains, indeed, the well-known law of conservation of electricity.\footnote{We may express our equations in the unified four-dimensional language of relativity with the usual notation:
\begin{equation*}\label{6a}
\begin{aligned}
   \frac{\partial F_{\nu i}}{\partial x_{\nu}}
 = \alpha^2 \varphi_i \qquad \qquad & (A) \\
F_{ik} = \frac{\partial \varphi_k}{\partial x_i}
       - \frac{\partial \varphi_i}{\partial x_k}  \quad &  (B)
\end{aligned}
\tag{6a}
\end{equation*}
From equations (\ref{6a}A) the equation of continuity of the vector potential follow:
\begin{equation*}\label{7a}
\frac{\partial \varphi_{\nu}}{\partial x_{\nu}} = 0 .
\tag{7a}
\end{equation*}
}
We do not want to conceal that in case we make everything real the general system of 2 times 8 equations, equivalent to the two equations of Dirac, contains more quantities and more equations than has been indicated so far. Maxwell's equations show a characteristic dichotomy to the effect that only an electric four-current occurs in them and no magnetic one. In the face of the extensive symmetry between $\mathcal{E}$ and $\mathcal{H}$ in the equations, this could not a priori be expected. Effectively, this dichotomy did not really come to light in the general scheme directly obtained by the transcription of the doubled Dirac equation. If there is also a magnetic current present, we may conceive the field as a superposition of two fields: the case with an electric current and without a magnetic one, on one hand, and, [on the other hand,] the opposite case with a magnetic current and without an electric one. We need only interchange the roles of $\mathcal{E}$ and $\mathcal{H}$ in the second case (to put $\mathcal{H}$ for $\mathcal{E}$  and $-\mathcal{E}$ for $\mathcal{H}$) and the field intensity can be expressed by a ``magnetic vector potential.'' Then the total $\mathcal{E}$ and $\mathcal{H}$ will be composed additively of the corresponding partial expressions. These make 6 equations. There is further the assertion that the electric current is proportional to the electric vector potential and the magnetic to the magnetic one; they make 2 times 4, i.e., 8 equations. Finally, the equation of continuity for the electric and magnetic currents are again 2 equations: altogether, there are 16 equations.

   Indeed, there here are two more scalars in the general equations of Dirac which Maxwell's theory completely lacks --- precisely as a consequence of the equation of continuity. If we retained them, (and thus dropped the equation of continuity), we would in full generality have the following quantities: electromagnetic field intensity (6 components), electric and magnetic vector potential (2 times 4, i.e., 8 components) and 2 scalars; altogether, 16 field quantities in accord with the 8 complex $\psi$ quantities. In the new interpretation, everything becomes real and only reasonable field theoretical quantities will play a role.

   Now we have important reasons to reduce the number of these quantities and remain at the already given equations \eqref{6} for which this reduction has already been performed.

   Let us consider from the point of view of our assignment, the four expressions~\eqref{2} bilinearly constructed from the fundamental quantities which according to Dirac form a four-vector: They have here a significantly different interpretation due to the modification of the transformation properties. They transform no longer as current densities and charge density (as a four-vector) but as a current of energy and an energy density. Dirac's conservation law turns simply into the conservation of energy.

   Actually, the remarkable circumstance holds that these expressions, when calculated with our field quantities, coincide completely with the Poynting vector and the energy density of Maxwell's theory, insofar as they depend on the electromagnetic field intensities. But additional terms still come in with the vector potential included. If we put $\alpha = 0$, then we should have Maxwell's equations in vacuum and, at the same time, the energy balance of Maxwell's theory in an unchanged form. As a consequence of the modification of Maxwell's equations by introducing terms containing $\alpha^2$ in system (\ref{6}A), the expressions for energy flux and energy density will also be modified. Additive terms appear with the factor $\alpha^2$, which now depend no more on the field intensity but on the vector potential, so that its character as a primary field quantity becomes manifest here, too.

   Quite similar circumstances prevail in the 3 equations, which can be added to the conservation of energy and which express the conservation of momentum.

   Altogether, the 16 quantities appearing in the conservation laws of momentum and energy form as a whole the so-called ``stress-energy tensor.'' A fundamental feature of this tensor is its symmetry. This important feature would be lost should we keep all the quantities which appear in the general scheme. In particular, the two scalars which are missing in Maxwell's theory determine an asymmetry which manifests itself by the presence of a magnetic current beside the electric. If we intend to retain the fundamental symmetry of the stress-energy tensor, we ought to drop all quantities which are also foreign to Maxwell's theory. At the same time, the dichotomy having come to light in Maxwell's equations would become well-understood.

   Although we started with a doubling of Dirac's equation and therefore also doubled the number of fundamental quantum mechanical quantities, we arrived by this additional restriction at a system which does not exceed by very much a single equation of Dirac (10 field quantities instead of 4 complex $\psi$). What is more, if we consider our system from the point of view of initial conditions, the restriction is even stronger than with Dirac. There is a time derivative in all the 4 functions there so that all the 4 equations may be freely chosen for the moment $t = 0$ in order to determine the system. Here, however, the total course in time will be determined if we prescribe $\mathcal{E}$ and $\mathcal{A}$ (i.e., only 6 conditions). Indeed $\varphi$ is obtained from the divergence of $\mathcal{E}$, the rotational of $\mathcal{A}$ determines $\mathcal{H}$, and the other field equations determine the temporal evolution of $\mathcal{E}$ and $\mathcal{A}$. (In Maxwell's equations, the freedom in the initial conditions is even smaller; $\mathcal{E}$ and $\mathcal{H}$ can be chosen freely but both are subjected to the condition of zero divergence.)

   It is very stimulating to see how both our field equations and our expressions for stress and energy are so closely related to Maxwell's theory and to the theory of electrons, respectively. If we could construct an electron from our equations, we should a priori be sure that the momentum-energy balance would hold at each point of the field. In this area, the theory of electrons is known to have grappled with great difficulties. To preserve this balance even at points where there is a charge, one had to resort to a hypothetical mechanical momentum and a corresponding energy. These factors emerge in our theory automatically in the correction terms mentioned above, which add to Maxwell's expressions and can effectively be interpreted as mechanical momentum and energy.\footnote{For details see ``The Conservation Laws in the Field Theoretical Exposition of Dirac's Theory,'' Zeits.\ f.\ Phys.\ {\bf 57}, 484, 1929. (\emph{Editorial note:} Ref.~\cite{LAN1929D}.)}

   In this circumstance, the idea suggests itself that a viewpoint could be found for the elimination of the conflict mentioned at the beginning between the field theoretical and quantum mechanical approaches and for the extension of the field theoretical basis to quantum phenomena. The equations of Dirac's theory we have dealt with so far hold only in the case of one electron, without external field. When an external electromagnetic field is present, Dirac's equations should be completed according to a certain prescription, in which the vector potential of the external field is introduced as a given quantity. If we consider the close relationship between the equations obtained and the theory of electrons, it must be highly improbable that a conventional theory of electrons might be constructed which gives us a vector potential, and then still another special theory of electrons for quantum mechanics. Because our theory already has a vector potential, it is very natural to use it also for the electromagnetic field of the electron, i.e., to identify it with the usual vector potential. The additional terms by which we have to complete the equations will then not postulate any external field quantities, but will contain only the proper field functions. In particular, they ought to occur \emph{quadratically}, because Dirac's additional terms are linear in the $\psi$ and linear in the external vector potential. The equations of the free electron thus would hold as a \emph{first approximation} to a system of equations that are nonlinear in reality, the contributions due to the external field representing second-order correction terms. Actually, we have ample indication that an adequate field theory can be built only on nonlinear equations. The fact that the solution of a linear equation remains a solution even when multiplied by an arbitrary factor indicates a freedom which has no counterpart in nature. It must, however, be added at once that this nonlinear extension of our equations, which should replace the introduction of external fields, has not yet been done in a satisfactory manner.

   We shall now consider our system \eqref{6} from the viewpoint that a fusion with the theory of electrons is conceivable in the sense that the classical theory of electrons would hold as a first approximation. It is easy to realize that this plan is not simple to implement. The constant $\alpha$ is not at all small but rather very large (of the order of $2.6 \times 10^{10}$ cm$^{-1}$) so that these terms in no case may be considered as a correction, nor can the normal vacuum equations of the theory of electrons be obtained even approximately. If one realizes that the quantum phenomena are actions at interatomic distances while the classical field theory governs the more distant regions void of matter, then one comes to suspect that the constant $\alpha$ is not to be regarded as a true constant. Rather, it is to be replaced by a field function that takes a high value in the central regions, but drops outwards to a very low value. In a physically consistent field theory it would be hopeless in any case to introduce a constant with the value of the mass of the electron: the mass difference between the electron and proton would be from the beginning deprived of an explanation.

   We can introduce another approach from a different viewpoint in order to intrinsically rehabilitate this assumption. Einstein's theory of relativity showed that energy is always linked with mass, and mass, for its part, also gravitates, i.e., the geometry of space is influenced. We might attempt to find a direct connection between the stress-energy tensor we found and the Riemannian curvature tensor as was the case with the stress-energy tensor of Maxwell's theory. However, since we are quite ignorant of the intrinsic meaning of our equations, a linkage without internal necessity would hardly lead to essentially new results, as was the case with Maxwell's theory.

   Instead, we shall admit another provisional consideration that is more rudimentary, but nevertheless plausible. The Lorentz transformation proves to be valid both for field theoretical and quantum mechanical phenomena. Granting this, the Lorentz transformation gives evidence of some important property of the geometry of space. In an arbitrarily curved space, a transformation of that kind, still anchored in Cartesian coordinates, makes no sense. Indeed, one could apply local Lorentz transformations but not universal ones, valid over all space. The mere fact of the existence of the Lorentz transformation seems to prove that the geometry could not strongly deviate from the Euclidean --- that the influence of the curvature has only a secondary effect, but does not come into question in first approximation.

   In reality, however, this is not the case. There is a Riemannian geometry to be considered as an immediate generalization of the Euclidean, which admits the Lorentz transformation just as the other. The line element of this geometry differs from the Euclidean one only by a common factor. We consider thus a geometry with the following structure:
\begin{equation*}\label{9}
ds^2 = \sigma (dx^2 + dy^2 + dz^2 + i^2 c^2 dt^2)
\tag{9}
\end{equation*}
or in a more condensed form:
\begin{equation*}\label{9a}
ds^2 = \sigma \, \sum dx_i^2 \, .
\tag{9a}
\end{equation*}
One is accustomed to saying: a geometry of this kind can be ``conformally'' mapped onto the Euclidean. The factor $\sigma$  that we shall for brevity call ``metric factor'' may be an arbitrary function of position (of the four coordinates).\footnote{This factor somewhat reminds us of the ``gauge factor'' of Weyl's theory but has a completely different sense because here we remain entirely within the framework of Riemannian geometry.}

   Every covariant equation in this geometry has an a priori structure that can also be interpreted as ``pure Euclidean,'' and is thus also consistent with the Lorentz transformation, though $\sigma$ may vary to any degree, and with it the geometry may even become approximately non-Euclidean. Of course, the function $\sigma$  will also occur in the equations. It is to be treated as a scalar under Lorentz transformation.

   Our equations \eqref{6} and \eqref{6a}, respectively, which presuppose an Euclidean metric, may easily be put in a generally covariant form, thereby rendering them compatible with a Riemannian basis field with arbitrary curvature. Let us take the geometry~\eqref{9} as a basis. It turns out that the system (\ref{6}B) that couples field intensity and vector potential remains unchanged. In system (\ref{6}A), however, the factor of measure $\sigma$ appears where we need it: in the mass term. When compared to the former case, the difference is only that in place of the constant $\alpha^2$ now $\alpha^2 \sigma$  is to be put on the right-hand side of system (\ref{6}A).

   In this way, we get a new possibility to introduce the mass term in the equations. Let us imagine we have set up our equations not with the very large constant $\alpha^2$ but with a very small one, which we shall denote by $k$, the value of which is still unknown. Furthermore, we shall consider the geometry modified in this specified way. It then happens that at larger distances, where the metric is practically Euclidean and the metric factor $\sigma$ drops to 1, the equations of the theory of electrons in vacuum turn out only slightly modified. In the central region of the electron, however, where a large amount of electric and mechanic energy is accumulated, the space will curve to such an extent that $\sigma$ reaches a very high value, whereby the term $k \sigma$ becomes very large, on the average comparable to the former $\alpha^2$ in Dirac's theory).\footnote{The equations \eqref{6} read in generally covariant form:
\begin{equation*}\label{10}
\begin{aligned}
\frac{1}{\sqrt{g}} \frac{\partial \sqrt{g} F^{\nu i}}{\partial x_{\nu}} = \alpha^2 \varphi^i \quad & (A) \\
F_{ik} = \frac{\partial \varphi_k}{\partial x_i} - \frac{\partial \varphi_i}{\partial x_k} \, . \quad & (B)
\end{aligned}
\tag{10}
\end{equation*}
With the accepted metric:
\begin{equation*}
F^{ik} = \frac{1}{\sigma^2}F_{ik} \, , \quad \sqrt{g} = \sigma^2 \, , \quad \varphi^i = \frac{1}{\sigma} \varphi_i \, ,
\notag
\end{equation*}
thus, the system (\ref{10}A) can be put in the form:
\begin{equation*}\label{11}
\frac{\partial F_{\nu i}}{\partial x_{\nu}} = k \sigma \varphi_i
\tag{11}
\end{equation*}
where $k$ figures in place of $\alpha^2$. The equation of continuity:
\begin{equation*}
\frac{1}{\sqrt{g}} \frac{\partial\sqrt{g} \, \varphi^{\nu}}{\partial x_{\nu}} =0
\notag
\end{equation*}
reads now:
\begin{equation*}\label{12}
\frac{\partial \, \sigma \varphi_{\nu}}{\partial x_{\nu}} = 0.
\tag{12}
\end{equation*}
}

   Is what we now have achieved to be regarded only as a formal improvement or is it also a deepening of the theory of electrons? May we hope to get nearer to the theory of matter in this way? It would be possible if we succeeded in constructing a formation from our field equations which would represent the electron.

   Apparently, the de Broglie's ``phase waves'' to which Dirac's theory is fitted, are scarcely suitable for giving a field theoretical representation of the electron. They describe oscillations in time, while there is scarcely any doubt that a really satisfactory field theory can be constructed only statically, in face of the high stability the electron shows externally.

\begin{figure}
\begin{center}
\resizebox{10cm}{!}{ \includegraphics{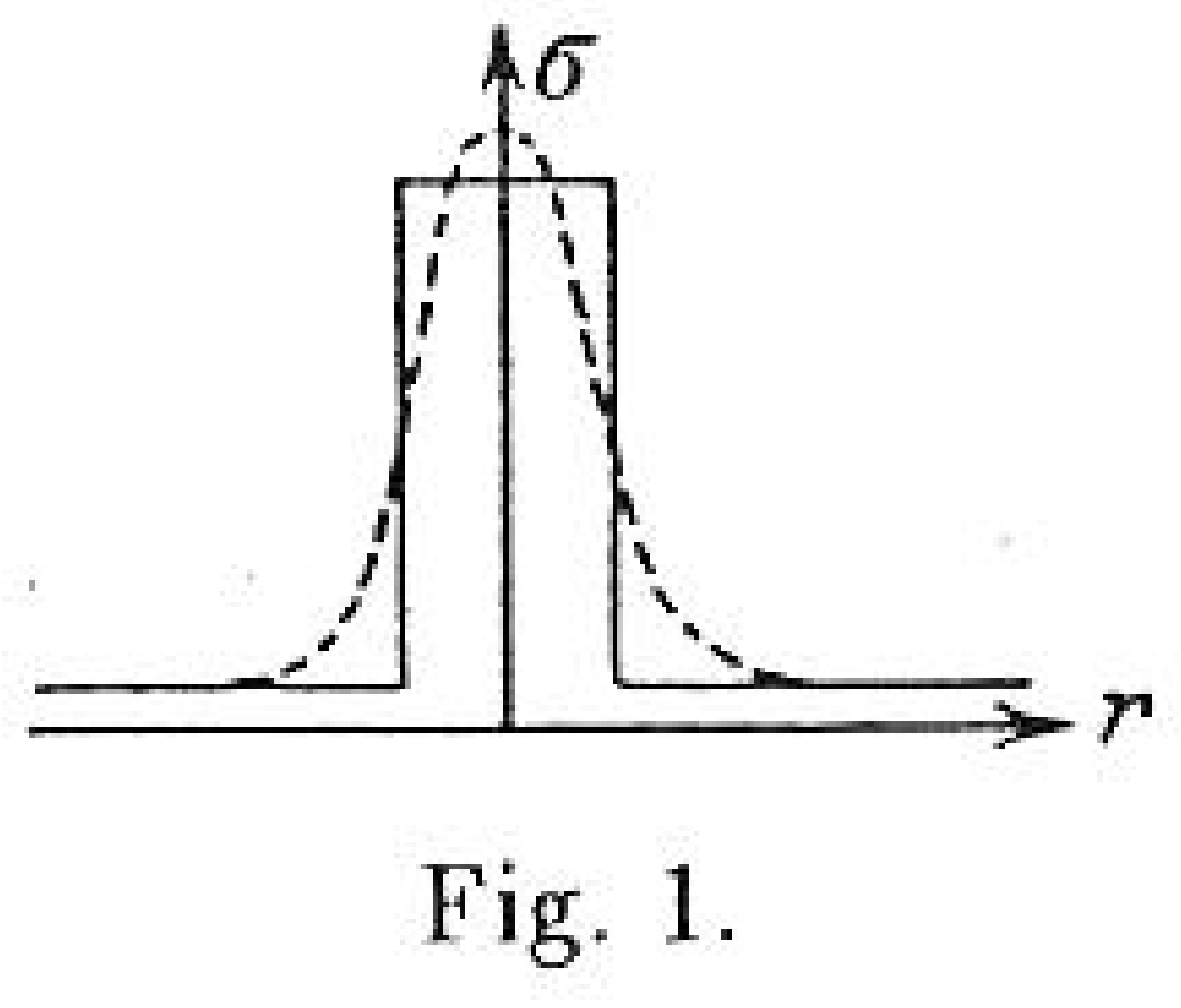}}
%---------------------------------------------
%\caption{ }
\end{center}
\end{figure}

   Do our field equations possess static solutions? What is more, are they everywhere regular solutions, in the sense of Schr\"odinger's ``eigenvalue physics,'' because only solutions of that kind could be seriously considered? There are, of course, solutions with singularities, but were we to permit them, our equations would not surpass the classical field equations in anything and should meet with the same difficulties they do.

   In the static case with spherical symmetry, the spatial vector potential must be $\mathcal{A}=0$. This we realize at once from the equation of continuity. The field will be determined by the scalar potential $\varphi$ alone. We have for it the equation:
\begin{equation*}\label{13}
\Delta \varphi = k \sigma \varphi
\tag{13}
\end{equation*}
($\Delta$ is the three-dimensional Laplacian operator) .

   This equation can under no conditions have non-zero regular solutions, as is easy to show by an integral conversion, as long as the right-hand side is everywhere positive. $k$ is a positive constant, as is $\sigma$ which by its meaning can never be zero or even have negative values. We must abandon our project, or declare it unworkable for the time being, until the equations get a more exact formulation, namely, by completing them with the quadratic terms already mentioned.

   It is nevertheless remarkable that the impossibility of static solutions is not a necessary consequence but only a consequence of the introduction of $k$ as a positive constant. It would not alter the general characteristics of the system of equations were $k$ negative and, though the idea is speculative, it seems to me not without interest to scrutinize this possibility. We should then really have the textbook example of a consistent field theory, capable of producing the electron as a regular, static formation.

   We ought to know, of course, about the field dependence of the geometric function $\sigma$. Because of our ignorance, we shall simplify the problem, which is here possible without drastically changing the general features of the solution.  How it is to be done can be seen in Fig.~1. The dotted curve denotes the suspected course of the function as a function of the distance $r$, the only variable of a solution with spherical symmetry. The cornered full curve is the schematic representation\footnote{This type of simplification has often been used in quantum mechanical investigations; e.g., the well-known study of Gamow (Zeits.\ f.\ Phys.\ {\bf 51}, 204, 1928) on radioactive decay.} which has a constant higher value within a certain region (``the electron radius'') then a sudden drop at the constant value~1 in the external space.

\begin{figure}
\begin{center}
\resizebox{8cm}{!}{ \includegraphics{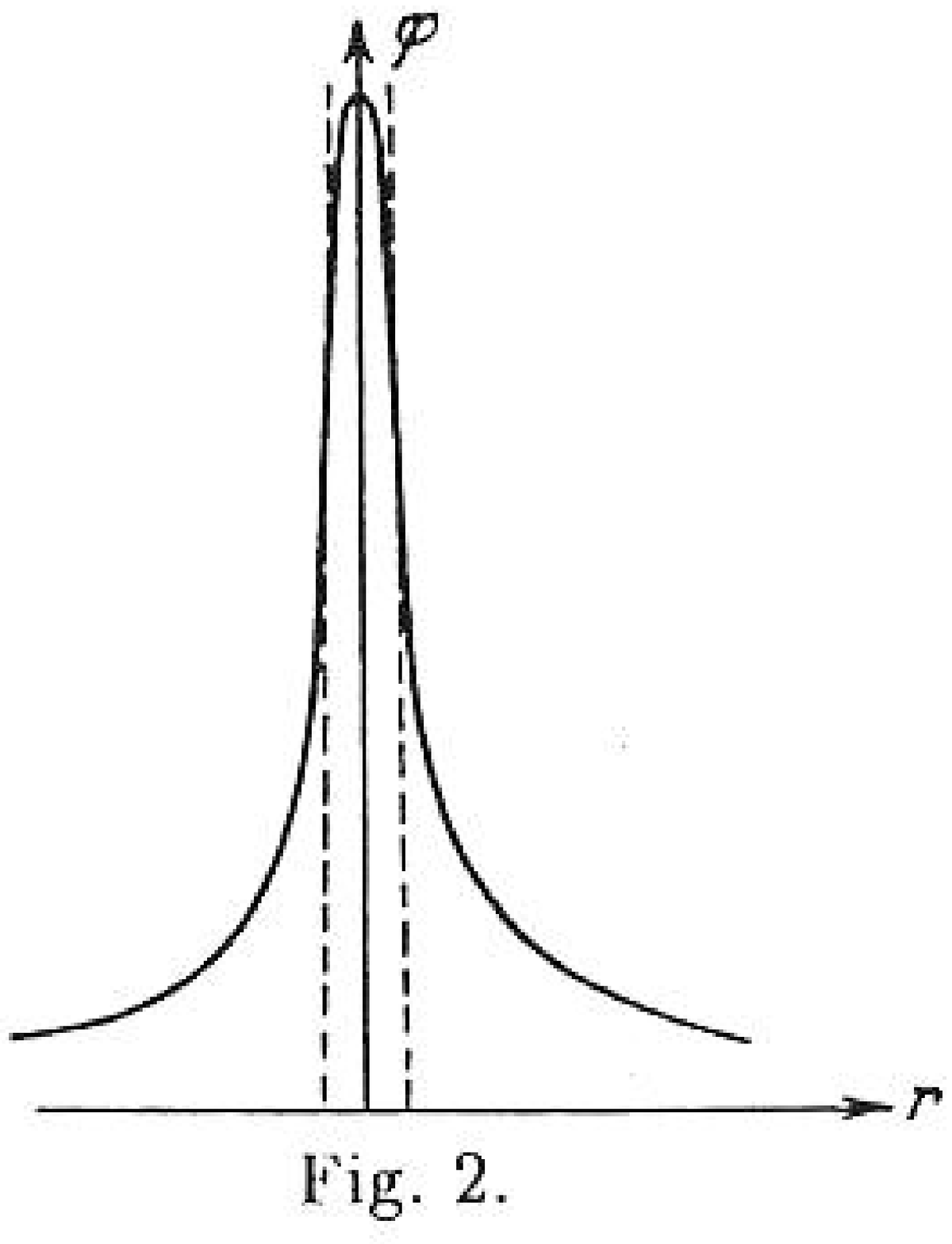}}
%---------------------------------------------
%\caption{ }
\end{center}
\end{figure}

   The corresponding curve of the electrostatic potential is given in Fig.~2. In the external space the classical ${1}/{r}$ law is valid, while in the internal space a ``rounding-off'' takes place with a continuous joining also in the tangents.

   If we widen the threshold for a given height (i.e., increase the ``radius''), we obtain Fig.~3. Outside, we find the law $\pm {1}/{r}$ again; inside, [we find] the law ${1}/{r} \sin ({2 \pi r}/{\lambda_0})$ with a continuous joining at the boundary which must lie at a distance of odd multiples of $\lambda_0$ from the center. The wavelength $\lambda_0$ may be calculated from the height of the threshold taken as the constant of Dirac's theory, as follows:
\begin{equation*}\label{14}
\frac{2 \pi}{\lambda_0} = \alpha = \frac{2 \pi m c}{h}
\tag{14}
\end{equation*}
consequently:
\begin{equation*}\label{15}
\lambda_0 = \frac{h}{m c}
\tag{15}
\end{equation*}

   Instead of de Broglie's characteristic frequency, a characteristic wavelength appears here. The phase wave of de Broglie, assigned to an electron at rest, consists of oscillations in time with spatially constant amplitude. The case inside the electron is the reverse: there are spatial vibrations in the spatial distribution of the potential with amplitudes constant in time. Space and time have, in a way, interchanged their roles.

\begin{figure}
\begin{center}
\resizebox{10cm}{!}{ \includegraphics{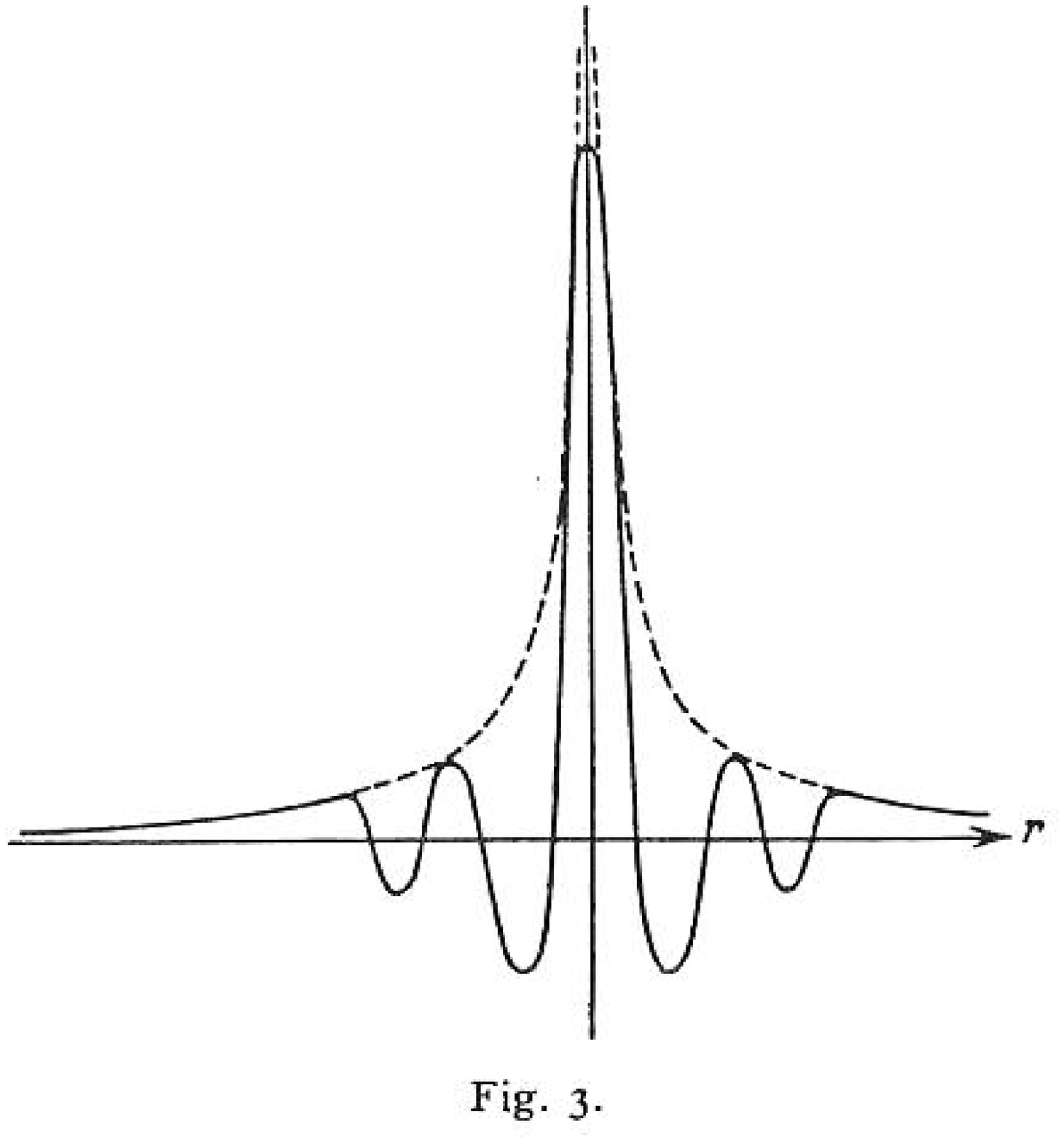}}
%---------------------------------------------
%\caption{ }
\end{center}
\end{figure}

  The equation being linear, a factor of proportion is left free in the solution, and that contradicts the definite value $e$ of the charge. We must recall, however, that in reality the equation is nonlinear, since $\sigma$ itself depends on the potential $\varphi$ in a still unknown way. Should we know this dependence, we could uniquely determine the height and width of the threshold, as well as the amplitude factor --- thus the charge of the electron. (Or rather 2 solutions ought to exist corresponding to both of the fundamental formations: the electron and proton.) We consider it as a favorable indication that in the height and width of the threshold two parameters are at our disposal, according to the two constants that macroscopically characterize the electron (and the proton, respectively): charge and mass.

   Let us consider a large number of electrons side by side on a plane, and sketch the behavior of the potential in this case by an infinite ``wall'' where we let the charges move.  We obtain what can be seen in Fig.~4: a linear rise outside, as in the classical case, while inside we have periodic vibrations with constant amplitude. If we put this wall in motion with velocity $v$, we need only apply a simple Lorentz transformation in order to obtain the change of the potential in the field. It follows for the internal periodic distribution:
\begin{equation*}\label{16}
\text{sin}  \frac{2\pi}{\lambda_0} \frac{z-vt}{\sqrt{1 - \frac{v^2}{c^2}}} .
\tag{16}
\end{equation*}
In a fixed point of space, an alternating field will develop with frequency:
\begin{equation*}\label{17}
\nu = \frac{1}{\lambda_0} \frac{v}{\sqrt{1 - \frac{v^2}{c^2}}} = \frac{m c}{h} \frac{v}{\sqrt{1 - \frac{v^2}{c^2}}}.
\tag{17}
\end{equation*}
If the wall passes over a lattice, the elements of the lattice will be excited to vibrate with this frequency just as if a light wave would have swept over it with a corresponding frequency. These elements must therefore emit aether radiation and give rise to interference.

\begin{figure}
\begin{center}
\resizebox{10cm}{!}{ \includegraphics{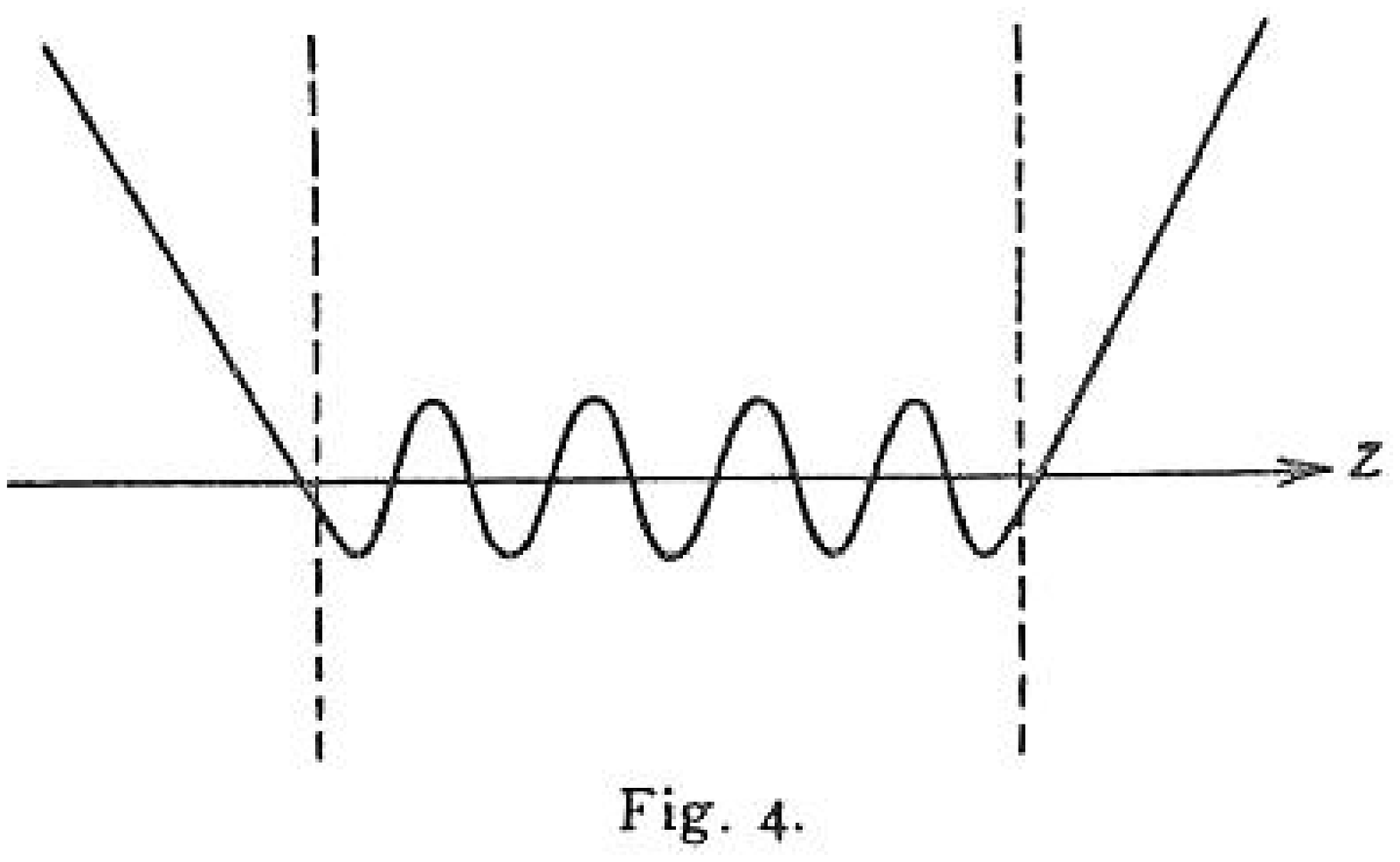}}
%---------------------------------------------
%\caption{ }
\end{center}
\end{figure}

   Equation \eqref{17} may also be put in the following form:
\begin{equation*}\label{18}
\frac{h \nu}{c} = \frac{mv}{\sqrt{1 - \frac{v^2}{c^2}}}.
\tag{18}
\end{equation*}
On the left there is the momentum of a light quantum, on the right the mechanical momentum of the electron, and we might say that a frequency is assigned to the moving electron stemming from the momentum relation, while with de Broglie it stems from the energy relation. Here space and time, in a way, again interchange their roles.

   The wavelength belonging to the aether radiation emitted by the lattice is:
\begin{equation*}\label{19}
\lambda = \frac{c}{\nu} = \lambda_0 \frac{c}{v}\sqrt{1 - \frac{v^2}{c^2}}.
\tag{19}
\end{equation*}
It is the wavelength experimentally found by Thomson, Davisson and Germer, Stern, Rupp, etc., in their well-known diffraction experiments.

   As suggestive as this deduction may be, it does not stand more severe tests. The idea is at variance with the usual view. Here we do not have those ghostlike phase waves --- that do not exist physically but are only to be ``assigned'' to the electron --- which interfere with one another and statistically prescribe the trajectory of the electron. Here the lattice is excited to emit real $\gamma$-rays. This realistic approach is, however, contradicted by experiment. What we really observe are not $\gamma$-rays but material radiation: deflected $\beta$-rays.

   It is still too early to tackle a problem as deep as the structure of the electron with these elementary assumptions [Ansätzen]. Nevertheless, I am of the belief that the investigation outlined above, which reveals formal relations of the Dirac theory to the equations of the electromagnetic field, goes beyond the purely formal to represent something deeper, though by far incomplete, toward an understanding of the quantum problems and the problem of matter, on a unified field theoretical basis.

\end{document}